\documentclass[twocolumn,pre,showpacs,floatfix]{revtex4}
\usepackage{graphicx}
\usepackage{amsfonts}
\usepackage{amssymb}
\usepackage{bm}

\newcommand{\be}{\begin{equation}}
\newcommand{\ee}{\end{equation}}
\newcommand{\bea}{\begin{eqnarray}}
\newcommand{\eea}{\end{eqnarray}}

\begin{document}

\title{Numerical Linked-Cluster Algorithms. \\ 
I. Spin systems on square, triangular, and kagom\'e lattices}

\author{Marcos Rigol}
\affiliation{Department of Physics and Astronomy, 
University of Southern California, Los Angeles, California 90089, USA}
\author{Tyler Bryant}
\affiliation{Department of Physics, University of California, Davis,
California 95616, USA}
\author{Rajiv R.~P.~Singh}
\affiliation{Department of Physics, University of California, Davis,
California 95616, USA}

\date{\today}


\begin{abstract}
We discuss recently introduced numerical linked-cluster (NLC) algorithms 
that allow one to obtain temperature-dependent properties of quantum lattice models, 
in the thermodynamic limit, from exact diagonalization of finite clusters. 
We present studies of thermodynamic observables for spin models on square, 
triangular, and kagom\'e lattices. Results for several choices of clusters
and extrapolations methods, that accelerate the convergence of NLC,
are presented. We also include a comparison of NLC results
with those obtained from exact analytical expressions (where available), 
high-temperature expansions (HTE), exact diagonalization (ED) of finite 
periodic systems, and quantum Monte Carlo simulations.
For many models and properties
NLC results are substantially more accurate than HTE and ED.
\end{abstract}

\maketitle

\section{Introduction \label{introduction}}

Accurate quantitative calculations of finite-temperature properties 
of quantum lattice models are a challenging task \cite{jaklic,sandvik}.
One of the few general methods that works directly in the thermodynamic limit,
is that of high-temperature expansions (HTEs), where properties of the system 
are expanded in powers of inverse temperature, $\beta=1/T$ \cite{domb}. 
These expansions, carried out to order $\beta^N$ (where $N$ is typically 
around 10), provide highly accurate numerical results at high temperatures. 
However, below some temperature $T_s$ related to the relevant microscopic 
energy scale of the system, the series diverges. Such a divergence
need not be related to any finite-temperature phase transitions or
long-range correlations. For example, in low dimensional or
frustrated spin systems, often there is either no finite-temperature phase
transition or such a transition occurs well below the exchange constant $J$.
The microscopic energy scale $J$ still controls the radius of convergence 
of the high-temperature series.

Beyond the radius of convergence $\beta>\beta_s$, series extrapolation 
methods \cite{guttmann} allow one to calculate thermodynamic properties, 
but their reliability remains uncertain. Our motivation for developing
the numerical linked cluster (NLC) method is to be able to obtain the 
properties of these systems, in the thermodynamic limit, for $\beta>\beta_s$ 
in a more reliable way, especially if the correlations in the system remain 
short ranged. NLC uses the linked cluster basis of HTE, but replaces the 
expansion in $\beta$ by an exact numerical calculation of the properties 
of linked clusters at any temperature \cite{rigol06,irving84}. In any 
practical implementation of NLC, only the contributions from clusters 
up to some maximum size are included. This can lead to highly accurate
properties of the thermodynamic system, even beyond the radius of
convergence of HTE, provided the correlations are short ranged. 
Thus, NLC helps to separate cases where the failure of HTE is due to its 
analytic structure in the complex plane, from where the correlations truly 
exceed the largest clusters studied. We would like to emphasize that this 
does not imply that comparable results for thermodynamic systems
can be obtained simply by exact diagonalization (ED) of individual 
periodic clusters of size comparable to the maximum size used in NLC.
We will show that NLC can be substantially more accurate than ED for 
finite-temperature properties of a thermodynamic system.
Furthermore, one can accelerate the convergence of NLC,
even when correlation length exceeds the largest cluster studied,
by using sequence extrapolation techniques, which are in
many ways analogous to series extrapolation methods \cite{guttmann,numrecipes}.

We present in this paper a detailed exposition of the NLC algorithm.
The basic method was already presented in Ref.\ \cite{rigol06}. 
Here we discuss in detail the different choices of clusters 
that can be used to build the numerical expansion. We also detail 
different extrapolation methods that, like Pade approximants for HTE, 
allow one to accelerate the convergence of NLC. These methods are especially 
relevant for the application of NLC to models in which correlations build 
up rapidly with reduction in temperature. Comparisons between results obtained 
by means of NLC with known techniques such as exact diagonalization (ED), 
quantum Monte Carlo (QMC) simulations, and HTE, are also presented.

The exposition is organized as follows. In Sec.\ \ref{basis}, we introduce 
the basis of NLC. We then present (Sec.\ \ref{extrapolation}) an overview
of different sequence extrapolation methods that can help accelerate the 
convergence of NLC. The rest of the manuscript is devoted to 
show how to build the series for spin models on square (Sec.\ \ref{SQ}), 
triangular (Sec.\ \ref{TR}), and kagom\'e (Sec.\ \ref{KA}) lattices. 
Different choices of clusters are discussed in detail, and applied to 
Ising, $XY$, and Heisenberg models. Finally, the conclusions are presented 
in Sec.\ \ref{conclusions}.

\section{Basis of NLC \label{basis}}

The fundamental basis for a linked cluster expansion, for some 
extensive property $P$ of an infinite lattice ${\cal L}$, 
is the relation \cite{domb,book}
\begin{equation}
P({\cal L})/N=\sum_c L(c)\times W_P(c),
\label{dirsum}
\end{equation}
where the left hand side is the value of the property $P$ per lattice site 
in the thermodynamic limit. On the right-hand side $L(c)$ is the so-called 
lattice constant, which is the number of embeddings of the cluster $c$, 
per lattice site, in the lattice ${\cal L}$ (explicit examples will be 
given later). $W_P(c)$ is the weight of the cluster $c$ for the property $P$. 
The latter is defined recursively by the principle of inclusion and 
exclusion \cite{domb},
\begin{equation}
W_P(c)=P(c)-\sum_{s\subset c}W_P(s),
\label{weights}
\end{equation}
where $P(c)$ is the property $P$ calculated 
for the finite cluster $c$. The sum on $s$ runs over all subclusters of $c$.
In HTE, for every cluster, $P(c)$ and hence its weight $W_P(c)$ are 
expanded in powers of $\beta$ and only a finite number of terms are retained. 
In NLC an exact diagonalization of the cluster is used to calculate 
$P(c)$ and $W_P(c)$ at any temperature. [Notice that in contrast 
to the clusters used in ED studies, the ones in Eq.\ (\ref{weights}) do 
not have periodic boundary conditions.] The exact numerical calculation of 
$P(c)$ allows NLC to build more bare information of the system 
than HTE.

There is another aspect in which the NLC scheme is fundamentally 
different from HTE, and that can be used to ones advantage. In HTE, 
the choice of clusters is dictated solely by the need to get the power
series expansion in $\beta$ to as high an order as possible.
This typically means that clusters are defined and ordered by
the number of bonds. In NLC, one has substantial freedom to select
the set of clusters and the order in which they are considered. One can arrange
them by number of sites, by number of bonds or, as we will see below,
one can even consider a reduced set of clusters. In order to obtain correct
results at high temperatures, one requirement is that with increasing 
order, the cluster weights, when expanded in inverse temperature, 
should give the correct HTE coefficients as well. 
This ensures that when HTE converges, NLC gives results that are
identical to HTE. However, as we will see, NLC expansions may 
involve a choice of clusters that sacrifice efficiency in the 
order to which they give the exact HTE coefficients, 
for better results at intermediate and low temperatures. 

As discussed before, when correlations are shorter ranged than the size of
clusters that are included in any implementation of NLC,
the NLC results are accurate without any need for extrapolations.
On the other hand, for systems with an ordered ground state, such an
implementation must eventually break down as the temperature is lowered 
(larger clusters will start making large contributions). 
In the next section we discuss some ``tricks'' that can be used to
accelerate the convergence of the direct sum in 
Eq.\ (\ref{dirsum}). This can lead to accurate thermodynamic
results at temperatures where correlation length far exceeds the sizes of the 
cluster included in the sum.

\section{Sequence extrapolations \label{extrapolation}}

In this section, we consider the general problem of estimating 
$P({\cal L})$ in Eq.\ (\ref{dirsum}), when the weights $W_P$ are 
only known for clusters up to a given size. In order to produce a 
sensible extrapolation scheme one can group clusters together and define
\begin{equation}
S_n=\sum_c L(c_n)\times W_P(c_n),
\end{equation}
where all clusters $c_n$ share a given characteristic, which could
be, for example, the number of bonds, sites, etc. 
Equation (\ref{dirsum}) can be rewritten as 
\begin{equation}
P({\cal L})=\sum_n S_n,
\label{partsum}
\end{equation}
and one can define partial sums as
\begin{equation}
P_n({\cal L})=\sum_{i=1}^n S_i.
\end{equation}
So, our goal is to estimate $P({\cal L})=\lim_{n\rightarrow \infty} P_n$ 
from a sequence $\lbrace P_n\rbrace$, which is known for $n=1,\ldots,N$.

Several methods have been developed to accelerate the convergence of such
sequences. An extensive review with references to original papers can be 
found in Ref.\ \cite{guttmann}, and they are similar to series
extrapolation methods. We will briefly describe here three methods 
that we have implemented, and that have proven to be particularly useful 
in accelerating the convergence of NLC. These methods are known as the Wynn's 
($\varepsilon$) algorithm \cite{guttmann}, the Brezinski's ($\theta$) 
algorithm \cite{guttmann}, and the Euler's transformation \cite{numrecipes}. 
A very important topic that is discussed neither here nor in 
Refs.\ \cite{guttmann,numrecipes} is one of error estimation. 
This is because most studies of errors associated with such extrapolation 
methods depend on the underlying function satisfying certain properties. 
As one might expect, these properties, in general, cannot be verified for many 
of the problems one finds in statistical mechanics.

Wynn's algorithm is defined as follows \footnote{Notice, that from the expression 
given in Ref.\ \cite{guttmann} we have corrected a sign in a subindex.}
\begin{eqnarray}
\varepsilon_n^{(-1)}&=&0,\qquad  \varepsilon_n^{(0)}=P_n, \nonumber \\
\varepsilon_n^{(k)}&=& \varepsilon_{n+1}^{(k-2)} + 
\frac{1}{\Delta\varepsilon_{n}^{(k-1)}},
\end{eqnarray}
where the discrete differentiating operator $\Delta$ is only applied to 
subscripts
\begin{equation}
\Delta\varepsilon_{n}^{(k-1)}=\varepsilon_{n+1}^{(k-1)} -\varepsilon_{n}^{(k-1)}.
\label{discreted}
\end{equation}
Within this scheme the even entries $\varepsilon_{n}^{(2k)}$ are expected 
to converge to $P({\cal L})$, while the odd ones 
$\varepsilon_{n}^{(2k+1)}$ diverge. Nonlinear sequence extrapolations 
usually display this behavior, and it implies that one has to be careful 
with round-off errors. 

One should notice that two iterations are needed for each level of
improvement so the new sequence is two terms shorter. We refer to each level 
of improvement as a cycle. For the problems we have studied so far by means 
of NLC, Wynn's algorithm has been, in general, the most successful in extending 
the region of convergency to lower temperatures.

Brezinski's algorithm is defined as follows:
\begin{eqnarray}
\theta_n^{(-1)}&=&0,\qquad  \theta_n^{(0)}=P_n, \nonumber \\
\theta_n^{(2k+1)}&=& \theta_{n}^{(2k-1)} + \frac{1}{\Delta\theta_{n}^{(2k)}},\nonumber \\
\theta_n^{(2k+2)}&=& \theta_{n+1}^{(2k)} + 
\frac{\Delta\theta_{n+1}^{(2k)}\Delta\theta_{n+1}^{(2k+1)}}{\Delta^2\theta_{n}^{(2k+1)}},
\label{Brez}
\end{eqnarray}
once again the discrete differentiating operator $\Delta$ is only applied to 
subscripts as in Eq.\ (\ref{discreted}) and
\begin{equation}
\Delta^2\theta_{n}^{(k)}=\theta_{n+2}^{(k)} - 2\theta_{n+1}^{(k)} + \theta_{n}^{(k)}.
\end{equation}

As for Wynn's algorithm, only even entries are expected to converge to
$P({\cal L})$. Once again, we have a cycle of improvement after two iterations, 
and for the Brezinski algorithm three terms are lost in each cycle. This fact, 
together with the second derivative in the denominator in $\theta_n^{(2k+2)}$ 
[Eq.\ (\ref{Brez})], reduce the number of cycles one can 
perform with the Brezinski algorithm as compared to Wynn's algorithm.

Finally, for alternating series, i.e., series in which terms $S_n$ 
[Eq.\ (\ref{partsum})] alternate in sign [$S_n=(-1)^n u_n$], 
there is a powerful tool known as the Euler transformation \cite{numrecipes}.
With this algorithm $P_{\infty}({\cal L})$ is approached by the sum 
($n$ is even)
\begin{equation}
u_0-u_1+u_2\ldots-u_{n-1}+\sum_s \frac{(-1)^s}{2^{s+1}} [\Delta^s u_n],
\label{Euler}
\end{equation}
where  $\Delta$ is the forward difference operator
\begin{eqnarray}
\Delta u_n&=&u_{n+1} - u_{n}, \nonumber\\
\Delta^2 u_n&=&u_{n+2}- 2u_{n+1} + u_{n}, \nonumber \\
\Delta^3 u_n&=&u_{n+3}- 3u_{n+2} + 3u_{n+1} - u_{n}, \ldots\ .
\end{eqnarray}
It is always advisable to do the sum of a small number of terms
directly, through term $n-1$, and then apply the transformation.

As we will show later, we have found Euler's transformation to be 
particularly useful for calculations of the specific heat. Having stated 
that NLC provides a scheme that similar to HTE allows for 
systematic extrapolations, we should add some remarks here. 
What the NLC scheme misses is the analytic structure of HTE.
These have proven particularly useful in studies of critical phenomena, 
where the region of diverging correlation length has also been addressed
using Pade extrapolations \cite{guttmann}. We have not yet investigated 
if NLC can be used to study critical phenomena, as our focus has been 
on models that do not order down to very low temperatures.

The analytic structure of HTE also allows for changes of variables, and
in some cases this can be used in very ingenious ways. For example, 
recent work by Bernu and 
Misguich \cite{bernu01}, has shown that by converting the expansion
for entropy in the inverse-temperature variable to an expansion
for entropy in the internal energy, one can develop a powerful 
extrapolation scheme that can build the ground-state energy and low-temperature
power-law behavior into the extrapolation of high-temperature expansions.
However, we note that such a scheme is only known for the entropy
(and related quantities) and not for susceptibilities or correlation functions.
We will show that for the specific heat (and entropy) of two-dimensional 
Heisenberg antiferromagnets the NLC scheme works better than direct 
extrapolation of HTE, but the method of Bernu and Misguich is better 
(at least for square and triangular lattice Heisenberg models) 
in that it allows estimates all the way down to $T=0$.

In what follows we study thermodynamic quantities (internal energy, entropy, 
specific heat, and uniform susceptibility) of spin models 
to show how different clusters and extrapolation techniques
can be used to build NLC on square, triangular, and kagom\'e
lattices.

\section{Square lattice \label{SQ}}

In this section we consider the square lattice. We discuss three 
different cluster schemes that we have used to build our NLC
expansions.

\begin{figure}[!htb]
\begin{center}
  \includegraphics[scale=.4,angle=0]{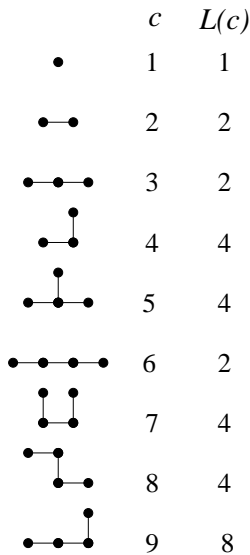}
\end{center}
\vspace{-0.5cm}
\caption{\label{BondGraphs}
All clusters with up to three bonds and their lattice constant 
for the square lattice.}
\end{figure}

The first, and most natural, choice is to consider all clusters 
and order them by the number of bonds. This selection has been 
called ``Weak Embeddings'' in the series expansion literature 
\cite{domb}, and is typically used for HTE. In Fig.\ \ref{BondGraphs}, 
we show all clusters that have up to three bonds, and their lattice 
constant.

Notice that one must include the single site cluster, which
corresponds to zero bonds. It dominates observables such as the entropy 
at very high temperatures. For the smallest bond clusters, such as the one
with one bond ($c=2$) and the first with two bonds ($c=3$), it is easy to 
realize that $L(c)=2$ since in the square lattice they can be only 
placed horizontally, and vertically. The second cluster with two bonds 
($c=4$) can be placed in four ways [$L(c)=4$], as one can realize rotating 
it by $90^\circ$, and so on.

\begin{table}[b]
\caption{Number of topological clusters and sum of the lattice constants
for clusters with up to 14 bonds in the square lattice. The cluster
with 0 bonds is the one site graph.}
\begin{tabular}{rrr}
\hline\hline
No.\ of bonds & No.\ of topological clusters & $\sum L(c)$ \\
\hline 
   0  &       1  &\qquad               1 \\
   1  &       1  &\qquad               2 \\
   2  &       1  &\qquad               6 \\
   3  &       2  &\qquad              22 \\
   4  &       4  &\qquad              88 \\
   5  &       6  &\qquad             372 \\
   6  &      14  &\qquad            1628 \\
   7  &      28  &\qquad            7312 \\
   8  &      68  &\qquad         33\,466 \\
   9  &     156  &\qquad        155\,446 \\
  10  &     399  &\qquad        730\,534 \\
  11  &    1012  &\qquad     3\,466\,170 \\
  12  &    2732  &\qquad    16\,576\,874 \\
  13  &    7385  &\qquad    79\,810\,756 \\
  14  & 20\,665  &\qquad 3\,86\,458\,826 \\
\hline\hline
\end{tabular}
\label{bondclusters}
\end{table}

It is convenient to group together all clusters that have the same
Hamiltonian, and diagonalize them just once. Since, the Hamiltonian
depends on the topology of how the sites are connected, we call
them topological clusters. Looking at our example,
clusters $c=3$ and 4, or $c=$\ 6--9 have the same 
topology. For calculating thermodynamic properties, all clusters
with the same topology make the same contribution. 
For the square lattice, we have calculated all possible clusters, and their 
lattice constants, up to 14 bonds. The number of topological clusters, 
and sum of $L(c)$, when grouped by their number of bonds is presented in 
Table.\ \ref{bondclusters}.

\begin{figure}[!htb]
\begin{center}
  \includegraphics[scale=.4,angle=0]{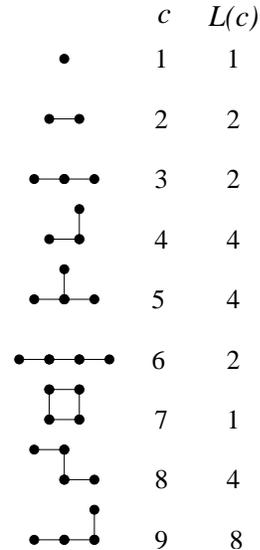}
\end{center}
\vspace{-0.5cm}
\caption{\label{SiteGraphs}
All clusters with up to four sites and their lattice constant 
for the square lattice.}
\end{figure}

A second choice is to identify clusters by sites. 
When building the Hamiltonian for such expansion, one 
places all possible bonds that connect any pair of sites in the graph. 
This leads to a set of clusters and embeddings that 
are called ``strong embeddings'' in the series expansion
literature \cite{domb}, and is typically used for generating 
the ``low-temperature expansions'' for Ising models. They can be 
used to generate HTEs as well. However, different clusters, with a given 
number of sites, will contribute to HTE in different orders. Thus, 
the order to which the HTE is correct will be determined by a subset of the 
clusters with the same number of sites that happen to contribute in the 
lowest order. On the other hand, having lots of compact clusters, with 
multiple connectivity between points, could mean that they can give better 
results beyond the radius of convergence of HTE. In Fig.\ \ref{SiteGraphs} 
we show all clusters that have up to four sites, and their lattice constant.

By comparing Figs.\ \ref{BondGraphs} and \ref{SiteGraphs} one can see
that the latter never includes graphs such as $c=7$ of the former one, i.e.,
all squares are always closed in the site expansion, hence the name
``strong embeddings''. In addition, while each site in the 
square lattice has four nearest-neighbor sites, each bond has six 
nearest-neighbor bonds, which implies that the number of bond clusters increases 
much faster than the number of site clusters. In the latter case, we have 
calculated all possible site graphs with up to 16 sites. Their number of 
topological clusters and sum of lattice constants, when grouped by 
number of sites, is shown in Table \ref{bondclusters}. 
\begin{table}[h]
\caption{Number of topological clusters and sum of lattice constants
for clusters with up to 16 sites in the square lattice.}
\begin{tabular}{rrr}
\hline\hline
No.\ of sites & No.\ of topological clusters & $\sum L(c)$ \\
\hline 
   1  &         1  &\qquad              1 \\
   2  &         1  &\qquad              2 \\
   3  &         1  &\qquad              6 \\
   4  &         3  &\qquad             19 \\
   5  &         4  &\qquad             63 \\
   6  &        10  &\qquad            216 \\
   7  &        19  &\qquad            760 \\
   8  &        51  &\qquad           2725 \\
   9  &       112  &\qquad           9910 \\
  10  &       300  &\qquad        36\,446 \\
  11  &       746  &\qquad       135\,268 \\
  12  &      2042  &\qquad       505\,861 \\
  13  &      5450  &\qquad    1\,903\,890 \\
  14  &   15\,197  &\qquad    7\,204\,874 \\
  15  &   42\,192  &\qquad   27\,394\,666 \\
  16  &  119\,561  &\qquad  104\,592\,937 \\
\hline\hline
\end{tabular}
\label{siteclusters}
\end{table}

Looking at Tables \ref{bondclusters} and \ref{siteclusters}
one can see that for NLC calculations of bond and site
based expansions the main limitation is the computing 
time (too many clusters) \footnote{This limitation, however,
can be mitigated by the fact that one can trivially 
parallelize the NLC calculation. Still, the number of clusters
increases exponentially so that only few orders can be gained 
by parallelization.}, and not the memory as is usual
for full diagonalization studies of clusters with periodic 
boundary conditions. Within NLC one can, however, change that 
order of limitations considering more complicated (larger) 
building units for the clusters. Hence, drastically reducing
the number of different clusters to consider \cite{rigol06}. 

\begin{figure}[!b]
\begin{center}
  \includegraphics[scale=.4,angle=0]{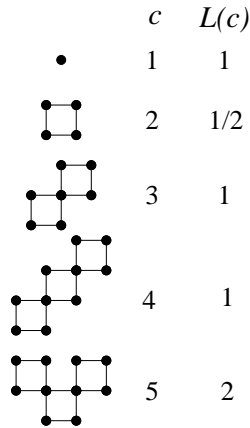}
\end{center}
\vspace{-0.5cm}
\caption{\label{SquareGraphs}
All topological clusters with up to three squares and their lattice constant 
for a square expansion.}
\end{figure}

A natural selection of a larger building unit in the square 
lattice is, of course, the elementary plaquette or square. 
In this case, 
a consistent NLC scheme requires that each bond belongs to 
only one square. This means that we build our clusters out
of every alternate square. Different squares can only share sites, 
which are the zeroth order of the square expansion, and are 
properly subtracted when calculating the weights in 
Eq.\ (\ref{weights}). In Fig.\ \ref{SquareGraphs} we show all 
clusters, with up to three such squares, required for a consistent 
square based NLC expansion.

The calculation of all possible clusters up to five squares (up to 16 sites) 
is in this case very simple. In Table \ref{squareclusters} we show 
the results for the number of topological clusters and sum of their lattice 
constants.
\begin{table}[h]
\caption{Number of topological clusters and sum of the lattice constants
for clusters with up to five squares in the square lattice. The cluster
with zero squares is the single site graph.}
\begin{tabular}{rrr}
\hline\hline
No.\ of squares & No.\ of topological clusters & $\sum L(c)$ \\
\hline 
   0  &   1  &\qquad     1 \\
   1  &   1  &\qquad   1/2 \\
   2  &   1  &\qquad     1 \\
   3  &   2  &\qquad     3 \\
   4  &   5  &\qquad  19/2 \\
   5  &  11  &\qquad  63/2 \\
\hline\hline
\end{tabular}
\label{squareclusters}
\end{table}

In the next subsections we apply the different NLC expansions detailed 
above to several well known spin models. All calculations were done on 
(3.2 GHz) Pentium IV personal computers in times that span between
16 h for the square based NLC expansion (up to 5 squares) 
and 60 h for the bond based NLC expansion (up to 13 bonds).

\subsection{Heisenberg model}

We now consider the antiferromagnetic Heisenberg model (AFHM) on 
the square lattice. Its Hamiltonian can be written as 
\begin{equation}
{\cal H}=\sum_{\langle i,j\rangle} {\bf S}_i\cdot {\bf S}_j,
\end{equation}
where we have chosen the coupling constant to be unity, 
and the sum runs over nearest-neighbor ($\langle i,j\rangle$) spins.

The AFHM on the square lattice is known to have an ordered ground state
with long-range antiferromagnetic correlations \cite{square-ref}.
This model can be efficiently simulated using
QMC techniques, such as stochastic series expansions \cite{sandvik91}. 
QMC methods enable one to study much larger system sizes than the 
ones accessible with exact diagonalization, although the classes of 
models that can be addressed are limited by the sign 
problem \cite{loh90,henelius00,troyer05}.

\begin{figure}[!b]
\begin{center}
  \includegraphics[scale=.6,angle=0]{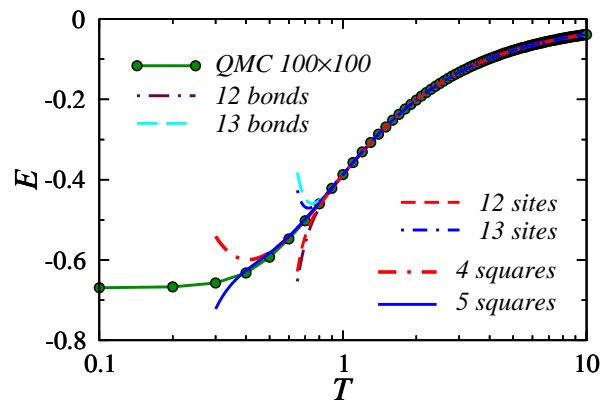}
\end{center}
\vspace{-0.5cm}
\caption{\label{BARE_EnergyHeisenberg}
(Color online) Energy as a function of temperature for the antiferromagnetic 
Heisenberg model on the square lattice. Bare NLC sums are compared with QMC 
results for a $100\times100$ lattice.}
\end{figure}

We start by studying the temperature dependence of the AFHM energy.
In Fig.\ \ref{BARE_EnergyHeisenberg}, we show a comparison of the 
bare sums for the bond, site, and square NLC expansions, with 
QMC results using the SSE technique \cite{yu06}. 
Several issues are apparent in Fig.\ \ref{BARE_EnergyHeisenberg}. 
(i) All NLC expansions give the correct result at high temperatures.
(ii) Direct NLC sums can converge below $T=1$, something that is not 
possible using HTE. (iii) The site expansion, which is closer in spirit 
to low-temperature expansions, performs better than the bond expansion
(closer to HTEs). This occurs even though the site expansion, up to the same 
order, is less demanding computationally than the bond expansion 
[there are many topological clusters in the bond expansion (1844) 
with 13 bonds and 14 sites, while in the site expansion only clusters 
with up to 13 sites were diagonalized].
(iv) The direct sum of the largest size cluster expansion, 
the square expansion in this case, converges to the lowest temperature 
($T\sim 0.5$), to be compared with ($T\sim 0.8$) for the site expansion.

\begin{figure}[!htb]
\begin{center}
  \includegraphics[scale=.6,angle=0]{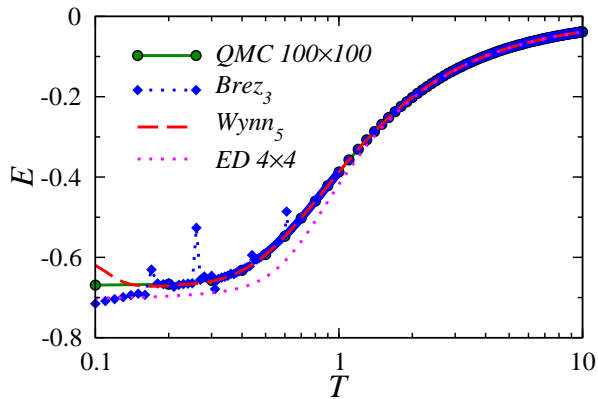}
\end{center}
\vspace{-0.5cm}
\caption{\label{EXTRAP_EnergyHeisenberg}
(Color online) Energy as a function of temperature for the antiferromagnetic 
Heisenberg model on the square lattice. Different extrapolations for the NLC site
expansion are compared with QMC results for a $100\times100$ lattice, and with 
exact diagonalization results for a $4\times4$ cluster (with periodic boundary
conditions).}
\end{figure}

The AFHM on the square lattice is known to develop
antiferromagnetic correlations at relatively high temperatures, i.e., 
it is the kind of model for which the direct NLC sum cannot converge up 
to very low temperatures. Now, the natural question that arises is how 
low in temperature can one go by using the sequence extrapolation
techniques discussed in Sec.\ \ref{extrapolation}. In Fig.\ 
\ref{EXTRAP_EnergyHeisenberg} we show two such extrapolations compared 
with the QMC results. The subindex following the name of each extrapolation
stands for the numbers of cycles of improvements done. In addition, 
for each cycle of improvement there are, in general, several terms. 
In what follows, when not explicitly specified otherwise, we will be 
showing the highest one.

Figure \ref{EXTRAP_EnergyHeisenberg} shows that extrapolations can indeed
work very well within NLC. We only show results for extrapolations of the site
expansion since for this expansion we obtain the best results. Already at
the level of the bare sum we saw that the site expansion performs better than 
the bond expansion. On the other hand, the square expansion, which produces
the best results for the direct sum, has too few terms to allow for a 
successful extrapolation scheme to work. [One can also realize that up to 
13 sites (with 8739 different topological clusters), the site expansion has explored 
much more of the lattice than the square expansion (with only 21 different 
topological clusters).] For the site expansion both Wynn's and Brezinski's 
algorithms converge, and agree with QMC, down to $T\sim 0.15$. One should 
notice, however, that while Wynn's results are smooth and on top of the QMC 
results all the way down to $T\sim 0.15$, some points in the Brezinski's 
extrapolation fall away from that curve. This is the kind of numerical 
problem one can run into after several orders of extrapolations. However, 
with the exception of these few points, Brezinski's results are still well 
converged and on top of the QMC data. Also shown is the exact diagonalization 
result for a $4\times 4$ cluster with periodic boundary conditions. It shows 
substantial finite-size effects already above $T>1$.

We now consider other thermodynamic quantities of interest, such as
the entropy and the specific heat. For the former, we have already shown 
\cite{rigol06} NLC results for Brezinski's and Wynn's extrapolations 
compared with the results of Bernu and Misguich \cite{bernu01} (obtained by 
integrating their specific heat curves). They exhibit a perfect agreement 
down to $T\sim 0.3$, where $S\sim 0.05$. Here, we will show the results 
for the specific heat.

\begin{figure}[!htb]
\begin{center}
  \includegraphics[scale=.6,angle=0]{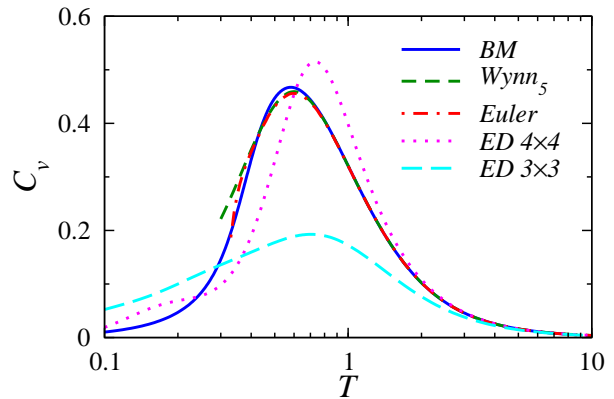}
\end{center}
\vspace{-0.5cm}
\caption{\label{EXTRAP_CvHeisenberg}
(Color online) Specific heat as a function of temperature for the 
antiferromagnetic Heisenberg model on the square lattice. Different 
extrapolations for the NLC site expansion are compared with the results 
of Bernu and Misguich \cite{bernu01} and with exact diagonalization results 
of small clusters (with periodic boundary conditions).}
\end{figure}

In Fig.\ \ref{EXTRAP_CvHeisenberg} we compare NLC results for the specific 
heat (after extrapolation) with the ones obtained by Bernu and 
Misguich \cite{bernu01}. Both approaches basically agree in the location 
of the specific heat peak, although they give slightly different 
peak values. Since NLC results have not converged fully below the peak, 
they may lead to the same results as Bernu and Misguich if a few more orders 
are included. Still, for $C_v$, NLC performs much better than direct Pade 
extrapolation of HTEs. QMC simulations for the specific heat also suffer from 
large errors ($C_v$ has to be obtained as derivative of the energy), and do 
not allow one to select any of the linked cluster results over the other \cite{yu06}. 

We also show in Fig.\ \ref{EXTRAP_CvHeisenberg} the specific heat 
results from the exact diagonalization of $4\times4$ and $3\times 3$ clusters
with periodic boundary conditions (PBCs), which helps to give an idea of 
the order of finite size effects in this model. They lead to a peak in 
the specific heat that is neither correct in its position nor height. 
There is another point to consider when comparing NLC with ED especially
in dimensions greater than 1. In the former case, we consider all possible 
clusters that make the NLC expansion consistent, without any biasing for 
any type of order. In exact diagonalization studies, PBCs
can bias the system towards or away from certain types of order.
For example, in the AFHM the $3\times3$ cluster with PBCs has much
bigger finite-size effects, because PBCs frustrates antiferromagnetism.
This issue may become important when the model under consideration has 
several competing orders, and the selection of a particular finite size 
cluster may artificially favor one order over the other. NLC, similar to 
HTE, does not suffer from this problem, and gives an unbiased answer for
the thermodynamic properties.

\subsection{$XY$ model in a staggered transverse field}

\begin{figure}[!b]
\begin{center}
  \includegraphics[scale=.6,angle=0]{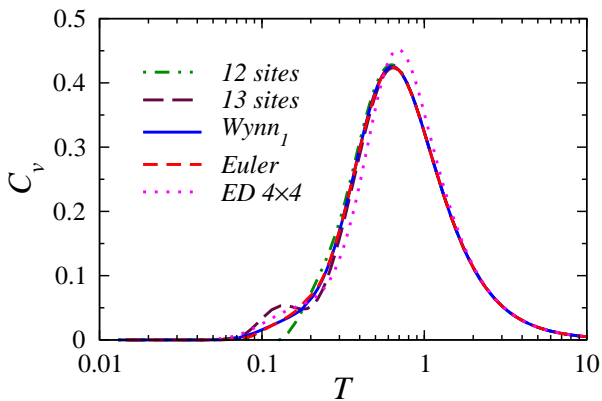}
\end{center}
\vspace{-0.5cm}
\caption{\label{EXTRAP_CvXY1.0}
(Color online) Specific heat as a function of temperature for the $XY$ model 
in a staggered transverse field. $\Delta=1$ so that the system is in the 
insulating regime. Direct sums and different extrapolations for the NLC 
site expansion are compared with exact diagonalization results 
for a $4\times4$ cluster (with periodic boundary conditions).}
\end{figure}

As discussed in Ref.\ \cite{rigol06}, NLC is ideal to study
models that stay short ranged at all temperatures or models in which
correlations build up slowly. In this section we discuss an example 
of the former case, the $XY$ model in a staggered transverse field.
Its Hamiltonian can be written as
\begin{equation}
{\cal H}=\sum_{\langle i,j\rangle} \left( S_i^x S_j^x + S_i^y S_j^y\right) 
+ \Delta\sum_i (-1)^{i_x+i_y} S_i^z,
\end{equation}
where we have chosen the $XY$ coupling to be unity, the sum runs 
over nearest-neighbor ($\langle i,j\rangle$) spins, and the last 
term describes the staggered field with strength $\Delta$.

The $XY$ model in a staggered transverse field can be mapped onto a 
hard-core boson model, at half filling, with a staggered site-dependent 
chemical potential. Such a model has been recently 
studied in one \cite{rousseau06}, two \cite{priyadarshee06}, and three 
dimensions \cite{aizenman04}. In 1D, due to its mapping to noninteracting
fermions, one can realize that there is, at zero temperature,
a phase transition from a superfluid to an insulating phase for $\Delta_c=0$,
i.e., any finite $\Delta$ produces an insulating phase \cite{rousseau06}. 
In two dimensions this model has been studied by means of QMC 
simulations, in this case the transition between the superfluid (also 
Bose-Einstein condensed phase) and the insulating phase occurs, at 
zero temperature, when $\Delta_c=0.992$ \cite{priyadarshee06}. Finally, 
in three dimensions this model has been used to rigorously prove
the existence of Bose-Einstein condensation and Mott insulating phases when
tuning the strength of the staggered chemical potential \cite{aizenman04}.

In what follows we consider the two-dimensional case.
In Fig.\ \ref{EXTRAP_CvXY1.0} we show the specific heat as a function 
of the temperature in a case where the system is insulating at zero
temperature. We have chosen $\Delta=1$, close to the critical value $\Delta_c$ 
for the superfluid-insulator transition. Due to the presence of a gap at 
zero temperature we can be certain that correlations stay short ranged 
at all temperatures. However, they are not negligible, 
and the direct sum of the NLC site expansion exhibits a small oscillation 
below the peak in the specific heat. A fully converged result, at all 
temperatures, can be obtained after just one cycle of improvement 
with Wynn's algorithm or using Euler's transformation. 
Figure \ref{EXTRAP_CvXY1.0} also shows that the exact diagonalization 
results for a $4\times4$ cluster with periodic boundary conditions still 
suffer from apparent finite size effects. 

\begin{figure}[!htb]
\begin{center}
  \includegraphics[scale=.6,angle=0]{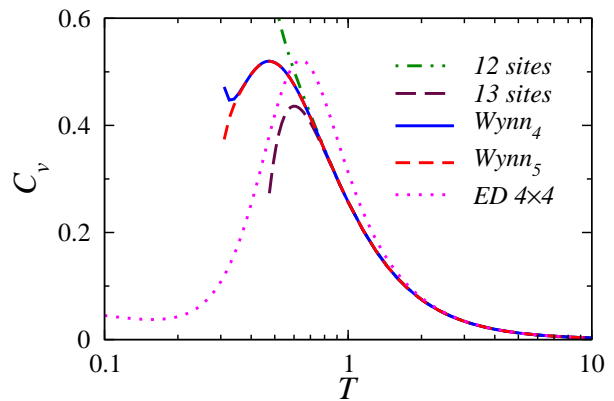}
\end{center}
\vspace{-0.5cm}
\caption{\label{EXTRAP_CvXY0.5}
(Color online) Specific heat as a function of temperature for the $XY$ model 
in a staggered transverse field, with $\Delta=0.5$. Direct sums and 
different extrapolations for the NLC site expansion are compared with
exact diagonalization results for a $4\times4$ cluster 
(with periodic boundary conditions).}
\end{figure}

Once in the regime of $\Delta$ where the system is superfluid at zero
temperature, the convergence of the NLC direct sum
does not reach very low temperatures, but sequence extrapolations 
allow one to reach the region below the peak in the specific
heat. This can be seen in Fig.\ \ref{EXTRAP_CvXY0.5}. As expected, 
in this case the difference with ED is even larger than when $\Delta>\Delta_c$.

\subsection{Ising model}

To conclude this section on spin models on the square lattice, we discuss
in this subsection the Ising model
\begin{equation}
{\cal H}=\sum_{\langle i,j\rangle} S_i^x S_j^x,
\label{isingH}
\end{equation}
which is an exactly soluble classical model \cite{ising}.

\begin{figure}[!htb]
\begin{center}
  \includegraphics[scale=.6,angle=0]{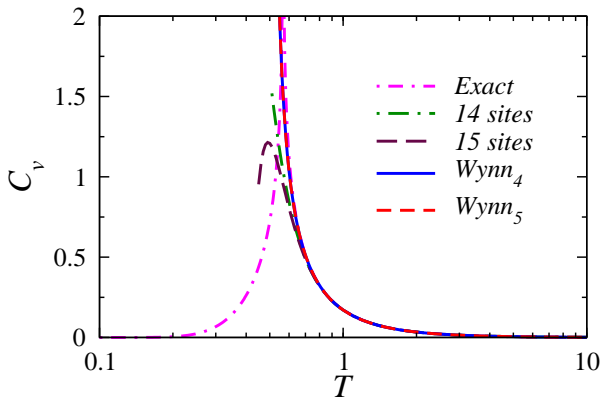}
\end{center}
\vspace{-0.5cm}
\caption{\label{EXTRAP_CvIS}
(Color online) Specific heat as a function of temperature for the Ising 
model. Direct sums and different extrapolations for the NLC site expansion 
are compared with exact analytical results.}
\end{figure}

In two dimensions, the Ising model is known to exhibit a finite-temperature 
transition between an ordered (gapped) phase, and a disordered 
high-temperature phase. As shown in Fig.\ \ref{EXTRAP_CvIS}, the specific
heat exhibits a divergence at the critical point, which is known to be 
logarithmic \cite{ising}. 
For NLC, the Ising model reduces to a counting problem as the Hamiltonian is
already diagonal. In Fig.\ \ref{EXTRAP_CvIS} we show direct sums for the
site based expansion up to 15 sites. Surprisingly, Wynn's extrapolations
allow one to obtain very good estimates of the specific heat very close to
the critical point.

We should stress, however, that for finite-temperature phase transitions 
with power-law singularities, HTE is probably the most efficient way to go. 
This is because such singularities can be built into the extrapolation.
We do not know if this is possible within NLC.

\section{Triangular lattice \label{TR}}

In this section we consider the triangular lattice. We discuss three 
different choices of basic clusters that allow one to build a consistent 
NLC expansion. 

The first possibility is, as in the square lattice, to build all 
possible clusters with up to a given number of bonds. Their number of 
topological clusters, and sum of $L(c)$, when grouped by the number of bonds 
is presented in Table \ref{bondclustersT}.
\begin{table}[t]
\caption{Number of topological clusters and sum of the lattice constants
for clusters with up to 12 bonds in the triangular lattice. 
The cluster with 0 bonds is the one site graph.}
\begin{tabular}{rrr}
\hline\hline
No.\ of bonds & No.\ of topological clusters & $\sum L(c)$ \\
\hline 
   0  &        1  &\qquad                1 \\
   1  &        1  &\qquad                3 \\
   2  &        1  &\qquad               15 \\
   3  &        3  &\qquad               91 \\
   4  &        5  &\qquad              603 \\
   5  &       12  &\qquad             4215 \\
   6  &       28  &\qquad          30\,535 \\
   7  &       72  &\qquad         226\,905 \\
   8  &      198  &\qquad      1\,718\,454 \\
   9  &      590  &\qquad     13\,207\,569 \\
  10  &     1817  &\qquad    102\,707\,301 \\
  11  &     5886  &\qquad    806\,366\,139 \\
  12  &  19\,753  &\qquad 2\,086\,381\,866 \\
\hline\hline
\end{tabular}
\label{bondclustersT}
\end{table}

From the number of topological clusters and the sum of lattice constants
in Table \ref{bondclustersT} one can see that the number of 
graphs in the triangular lattice grows much faster than in the 
square lattice. This is because in the triangular lattice each 
bond has ten nearest-neighbor bonds, as opposed to six in the 
square lattice. (The number of nearest-neighbor bonds determines 
the rate of growth of the number of possible clusters.)

\begin{table}[b]
\caption{Number of topological clusters and sum of the lattice constants
for clusters with up to 13 sites in the triangular lattice.}
\begin{tabular}{rrr}
\hline\hline
No.\ of sites & No.\ of topological clusters & $\sum L(c)$ \\
\hline 
   1  &       1  &\qquad             1 \\
   2  &       1  &\qquad             3 \\
   3  &       2  &\qquad            11 \\
   4  &       4  &\qquad            44 \\
   5  &       8  &\qquad           186 \\
   6  &      22  &\qquad           814 \\
   7  &      54  &\qquad          3652 \\
   8  &     156  &\qquad       16\,689 \\
   9  &     457  &\qquad       77\,359 \\
  10  &    1424  &\qquad      362\,671 \\
  11  &    4505  &\qquad   1\,716\,033 \\ 
  12  & 14\,791  &\qquad   8\,182\,213 \\ 
  13  & 49\,138  &\qquad  39\,267\,086 \\
\hline\hline
\end{tabular}
\label{siteclustersT}
\end{table}

A second natural choice is to build all clusters with up to a 
given number of sites. Once again, when building the Hamiltonian 
for such clusters one needs to place the maximum number of bonds 
possible in them. This selection of building blocks for the 
NLC expansion provides a significant reduction in the number of 
clusters one needs to consider as compared to the bond expansion.
(Each site in the triangular lattice has only six nearest neighbor 
sites.) In addition, having more compact clusters,
this expansion performs better and allows for better extrapolations 
in the intermediate and low-temperature regimes. The number of topological 
clusters and sum of lattice constants for the site
based NLC expansion is shown in Table \ref{siteclustersT}.

A triangular lattice is made out of triangles, so it is also possible
to develop a NLC for the triangular lattice where the clusters
consist of closed triangles. However, in this case, a consistent
NLC scheme requires that one restricts the calculation to a single site
plus clusters of up (or down) pointing triangles only. The reason for 
this restriction is that all bonds of the triangular lattice belong to 
a unique up (or down) pointing triangle. Different triangles should only 
share sites. The number of topologically distinct clusters with 1 
through 8 triangles, and the sum of their lattice constant is shown in 
Table \ref{triangleclusters}. (We have grouped them by the number of 
triangles.)
\begin{table}[h]
\caption{Triangular lattice number of topological clusters and sum of the 
lattice constants for clusters with up to eight triangles.
The cluster with 0 triangles is the single site.}
\begin{tabular}{rrr}
\hline\hline
No.\ of triangles & No.\ of topological clusters & $\sum L(c)$ \\
\hline 
   0  &       1  &\qquad          1 \\
   1  &       1  &\qquad        1/3 \\
   2  &       1  &\qquad          1 \\
   3  &       3  &\qquad       11/3 \\
   4  &       5  &\qquad       44/3 \\
   5  &      12  &\qquad         62 \\
   6  &      35  &\qquad      814/3 \\
   7  &      98  &\qquad     3652/3 \\
   8  &     299  &\qquad       5563 \\
\hline\hline
\end{tabular}
\label{triangleclusters}
\end{table}

Notice that one advantage of the triangle-based expansion in the
triangular lattice, over the square-based expansion in the square 
lattice, is that in the former the maximum number of lattice 
sites of a cluster with $N_t$ triangles is $2N_t+1$, while in the 
square lattice it is $3N_s+1$ ($N_s$ being the number of squares). 
This means that one can fully diagonalize clusters with more triangles 
than squares, which helps both for the bare NLC sums as well as for 
extrapolations. In the next subsections we apply the above expansions 
to Heisenberg and Ising models on the triangular lattice.

\subsection{Heisenberg model}

The triangular-lattice antiferromagnetic Heisenberg model is a 
fascinating quantum spin model, which has long-range order at
$T=0$ \cite{lhuillier} but with spin-spin correlations that
remain short range down to fairly low temperatures \cite{elstner93}. 
In contrast to the square lattice AFHM, the AFHM on the triangular 
lattice shows no evidence for a renormalized classical 
behavior \cite{chakravarty88,azaria92,chubukov94} 
down to lowest temperatures that can be reached in HTE.
It is a frustrated spin model so that QMC methods suffer from a 
sign problem. It has recently been argued \cite{zheng06,starykh02} 
that the anomalous finite-temperature behavior is due to the 
excitation of rotons, which leads to high entropy at relatively 
low temperatures.

The specific heat of the triangular lattice AFHM is a quantity that could be
of direct experimental interest. It has also been calculated from HTE
by the recent approach of Bernu and Misguich (BM) \cite{bernu01}. 
These authors found that direct Pade approximants \cite{elstner93}
not only fail at surprisingly high temperatures but are also unable to 
correctly locate the maximum of the specific heat and its height.

\begin{figure}[!htb]
\begin{center}
  \includegraphics[scale=.6,angle=0]{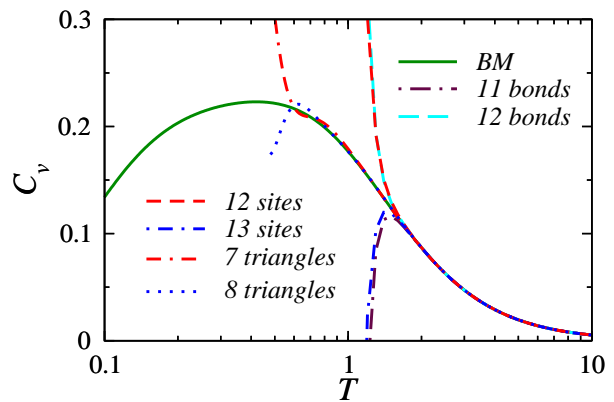}
\end{center}
\vspace{-0.5cm}
\caption{\label{BARE_CvHeisenbergTR}
(Color online) 
Specific heat as a function of temperature for the Heisenberg model 
on a triangular lattice. Direct sums are compared with BM results
in Ref.\ \cite{bernu01}.}
\end{figure}

In Fig.\ \ref{BARE_CvHeisenbergTR} we show the bare sums for the three
possible NLC expansions discussed before, and compare them with the BM 
results \cite{bernu01}. The bond and site expansions, up to 12 bonds and
13 sites, respectively, are well converged only at high temperatures 
(up to $T\sim 2$), with the site expansion being slightly better. On the 
other hand, the triangle based expansion converges down to a lower 
temperature $T\sim 0.6$. This temperature is very close to the lowest 
temperature up to which direct Pade extrapolations agree with BM results.

\begin{figure}[!htb]
\begin{center}
  \includegraphics[scale=.6,angle=0]{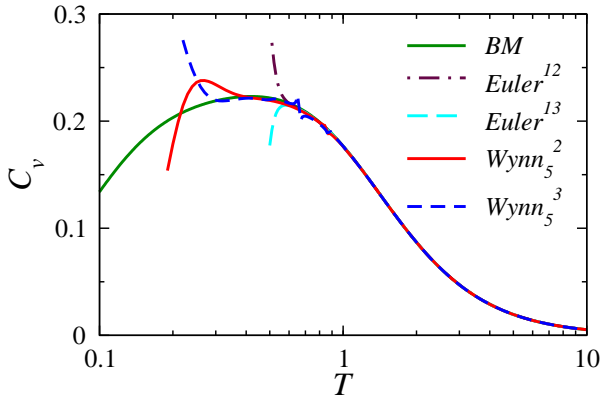}
\end{center}
\vspace{-0.5cm}
\caption{\label{EXTRAP_CvHeisenbergTR}
(Color online) 
Specific heat as a function of temperature for the Heisenberg model 
on a triangular lattice. Extrapolations are compared with BM results
in Ref.\ \cite{bernu01}. The superindex refers to the term in
the extrapolation (see text for details).}
\end{figure}

On performing extrapolations over the bare sums shown in Fig.\ 
\ref{BARE_CvHeisenbergTR}, we found that the site expansion 
is the one that enables an improvement of the convergence 
to the lowest temperature. (The bare results for the triangle based 
expansion can hardly be improved by sequence extrapolation methods.) 
In Fig.\ \ref{EXTRAP_CvHeisenbergTR} we show results for two possible
extrapolations of the site based NLC expansion compared with 
BM results \cite{bernu01}.

Notice that we have included two terms for each extrapolation scheme. 
To understand what they mean one has to realize that up to 13 sites 
Euler transformation allows for 13 terms, from which we have taken 
the first four to be the bare sums, and starting with the fifth we have 
applied the transformation as explained in Sec.\ \ref{extrapolation}. 
Hence, in Fig.\ \ref{EXTRAP_CvHeisenbergTR} we are showing the last 
two terms (in Sec.\ \ref{SQ} we showed only the last one). For Wynn's 
approach on the other hand, one should remember that two terms are lost 
after each cycle of improvement. So after, five cycles (the case in 
Fig.\ \ref{EXTRAP_CvHeisenbergTR}) ten out of the initial 13 terms in the 
bare sum are lost, i.e., in Fig.\ \ref{EXTRAP_CvHeisenbergTR} we are 
showing the last two of the remaining three.

Figure \ref{EXTRAP_CvHeisenbergTR} shows that while Euler transformation
allows one to extend the convergence of the site based expansion to the 
region where the triangle based expansion was convergent, Wynn's 
extrapolation scheme enables one to obtain results at lower temperatures 
(up to $T\sim 0.3$). In contrast to direct Pade for HTE, Wynn's scheme 
for NLC allows one to reach the maximum of the specific heat as predicted 
by BM \cite{bernu01}.

\begin{figure}[!htb]
\begin{center}
  \includegraphics[scale=.6,angle=0]{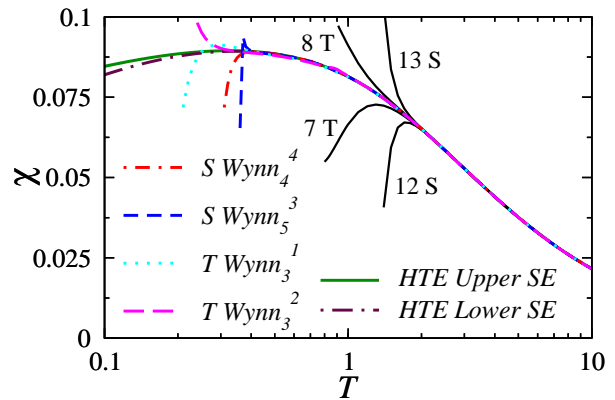}
\end{center}
\vspace{-0.5cm}
\caption{\label{EXTRAP_SusceptibilityHeisenbergTR}
(Color online) 
Uniform susceptibility as a function of temperature for the AFHM on a 
triangular lattice. NLC bare sums are shown for up to 7 and 8 triangles 
(7 T and 8 T in the figure), and up to 12 and 13 sites (12 S and 13 S 
in the figure). The corresponding extrapolations for the site based 
($S$ Wynn) and triangle based ($T$ Wynn) expansions are compared 
with series extrapolation results of HTE, obtained by integrated 
differential approximants \cite{coldea}. For the latter, only the 
upper and lower boundaries are shown.}
\end{figure}

A second quantity of much experimental interest
is the uniform susceptibility \cite{coldea}. 
In Fig.\ \ref{EXTRAP_SusceptibilityHeisenbergTR} we show NLC (bare and
extrapolated) results for the uniform susceptibility ($\chi$) of the 
AFHM on the triangular lattice. NLC results are compared for this 
quantity with series extrapolations of HTE obtained by integrated 
differential approximants \cite{coldea}. (Notice that the BM approach 
\cite{bernu01} is not suitable for calculations of $\chi$.)

A comparison between the results for $\chi$ and $C_v$ helps to understand
why the flexibility one has for selecting different kind of clusters in
building the NLC expansion can be useful. In contrast to $C_v$, the 
results of the bare sums for the site based and triangle based NLC 
expansions for the susceptibility
converge well up to about the same high temperature 
(see Fig.\ \ref{EXTRAP_SusceptibilityHeisenbergTR} vs 
Fig.\ \ref{BARE_CvHeisenbergTR}). This might suggest that the triangle 
based expansion may not bring any advantage for this quantity.
However, in contrast to $C_v$, series extrapolations extend the region 
of convergence for $\chi$ for the triangle based NLC ($T$ Wynn in 
Fig.\ \ref{EXTRAP_SusceptibilityHeisenbergTR}), and allow one to reach 
lower temperatures than the extrapolations for the site based expansion 
($S$ Wynn in Fig.\ \ref{EXTRAP_SusceptibilityHeisenbergTR}). Notice 
that in the region where NLC extrapolations are well converged 
they are in excellent agreement with extrapolations of HTE obtained 
by integrated differential approximants \cite{coldea}.
For $\chi$, the integrated differential approximants 
of HTE (which do not work so well for $C_v$) 
appear to have convergence to lower temperatures 
than the ones reached with NLC.

\subsection{Ising model}

The Ising model [Eq.\ (\ref{isingH})] on the triangular lattice is
an exactly soluble model \cite{wannier50,houtappel50,stephenson64}.
At zero temperature it exhibits power-law decaying spin correlations 
and an extensive entropy $S=0.3231$. 

\begin{figure}[!htb]
\begin{center}
  \includegraphics[scale=.6,angle=0]{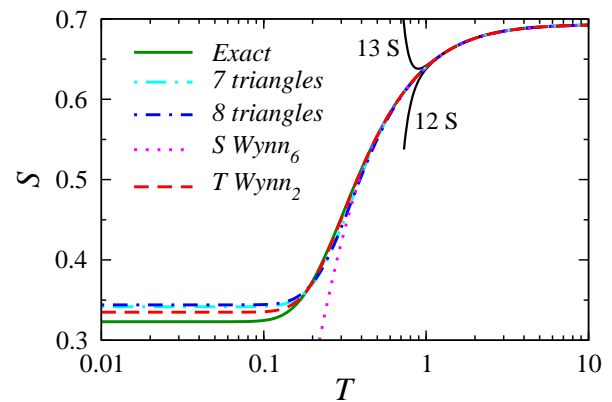}
\end{center}
\vspace{-0.5cm}
\caption{\label{EXTRAP_EntropyIsingTR}
(Color online) 
Entropy as a function of temperature for the Ising model on a 
triangular lattice. NLC bare sums are shown for up to 7 and 8 
triangles, and up to 12 and 13 sites (12 S and 13 S in the figure). 
The corresponding extrapolations for the site based (S Wynn) and 
triangle based (T Wynn) expansions are compared 
with the exact analytical result.}
\end{figure}

For this model, the triangle based NLC expansion converges down to 
low enough temperatures that it even allows one to study some ground 
state properties, for example, the entropy shown in 
Fig.\ \ref{EXTRAP_EntropyIsingTR}. One can see
that the site based expansion converges only up to $T\sim 1$. 
Hence, the triangle based expansion provides one with a qualitative 
improvement over site (and bond) expansion. Adding up contributions
from clusters up to $N_t=1$, 2, 3, 4, 5, 6, 7, 8 triangles leads to 
estimates of zero temperature entropy of $S=0.6931$, 0.4055, 0.3677, 
0.3677, 0.3499, 0.3521, 0.3417, 0.3440, respectively. In this case, 
the convergence to the thermodynamic limit result is power law in $1/N_t$, 
compared to an exponential convergence in the kagom\'e case \cite{rigol06}, 
which is not surprising given that the triangular-lattice model is critical 
\cite{stephenson64}.

Wynn's extrapolation of the triangle expansion, up to 8 triangles, 
improves towards the thermodynamic limit result as shown in 
Fig.\ \ref{EXTRAP_EntropyIsingTR}. At low temperatures it gives an estimate 
of the entropy $S=0.3349$. The extrapolation for the site expansion,
on the other hand, only converges down to $T\sim 0.3$.

\begin{figure}[!htb]
\begin{center}
  \includegraphics[scale=.6,angle=0]{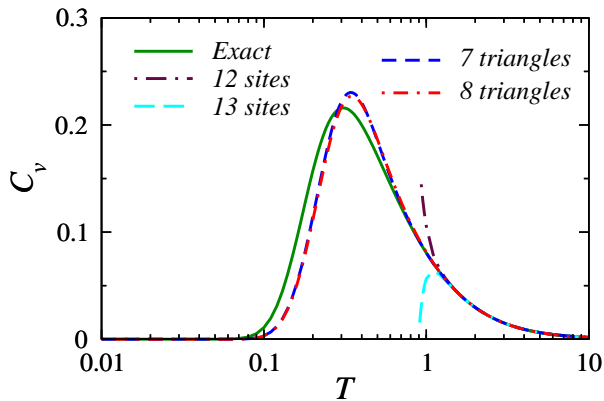}
\end{center}
\vspace{-0.5cm}
\caption{\label{BARE_CvIsingTR}
(Color online) 
Specific heat as a function of temperature for the Ising model on a 
triangular lattice. NLC bare sums are shown for up to 7 and 8 
triangles, up to 12 and 13 sites, and compared 
with the exact analytical result.}
\end{figure}

A comparison of the NLC results for the specific heat with the 
exact analytical calculation is shown in Fig.\ \ref{BARE_CvIsingTR}.
For this quantity, extrapolations of the site, bond, and triangle
based expansions do not allow one to improve over the direct triangle
based sum, so we do not show them there. It is remarkable, however, 
that even though the system is critical the results of the triangle based 
bare sums are not far from that exact result.

\section{Kagom\'e lattice \label{KA}}

In this section we consider the kagom\'e lattice. As before, we discuss 
three different choices of basic clusters that allow one to build a 
consistent NLC expansion. 

The first choice, again, is to use all clusters 
up to a given number of bonds. The number of topological clusters, 
and sum of $L(c)$, when grouped by their number of bonds 
is presented in Table.\ \ref{bondclustersK}.
\begin{table}[t]
\caption{Number of topological clusters and sum of the lattice constants
for clusters with up to 13 bonds in the kagom\'e lattice. 
The cluster with 0 bonds is the one site graph.}
\begin{tabular}{rrr}
\hline\hline
No.\ of bonds & No.\ of topological clusters & $\sum L(c)$ \\
\hline 
 0  &           1  &\qquad                 1   \\
 1  &           1  &\qquad                 2   \\
 2  &           1  &\qquad                 6   \\
 3  &           3  &\qquad                62/3 \\
 4  &           4  &\qquad                77   \\
 5  &           8  &\qquad               304   \\
 6  &          17  &\qquad              3752/3 \\
 7  &          36  &\qquad              5294   \\
 8  &          81  &\qquad           22\,845   \\
 9  &         194  &\qquad          299\,924/3 \\
10  &         481  &\qquad          442\,507   \\
11  &        1235  &\qquad       1\,977\,572   \\
12  &        3297  &\qquad      26\,729\,935/3 \\
13  &        8944  &\qquad      40\,418\,174   \\
\hline\hline
\end{tabular}
\label{bondclustersK}
\end{table}

A second choice is to build all clusters with up to a given 
number of sites. When building the Hamiltonian for such clusters 
one needs to place the maximum number of bonds possible in them. 
The number of topologies and sum of lattice constants for the site
based NLC expansion is shown in Table \ref{siteclustersK}. We have 
grouped them by number the number of sites.
\begin{table}[h]
\caption{Number of topological clusters and sum of the lattice constants
for clusters with up to 15 sites in the kagom\'e lattice.}
\begin{tabular}{rrr}
\hline\hline
No.\ of sites & No.\ of topological clusters & $\sum L(c)$ \\
\hline 
 1  &           1  &\qquad                 1   \\
 2  &           1  &\qquad                 2   \\
 3  &           2  &\qquad                14/3 \\
 4  &           2  &\qquad                12   \\
 5  &           4  &\qquad                33   \\
 6  &           7  &\qquad               281/3 \\
 7  &          12  &\qquad               272   \\
 8  &          22  &\qquad               805   \\
 9  &          45  &\qquad              2420   \\
10  &          88  &\qquad              7358   \\
11  &         183  &\qquad           22\,581   \\
12  &         389  &\qquad          209\,552/3 \\
13  &         842  &\qquad          217\,522   \\
14  &        1855  &\qquad          681\,224   \\
15  &        4162  &\qquad       2\,143\,905   \\
\hline\hline
\end{tabular}
\label{siteclustersK}
\end{table}

Since the kagom\'e lattice consists of corner sharing triangles,
the triangle-based NLC, in this case, involves all elementary
triangles. This selection of building blocks for NLC reduces dramatically 
the number of clusters to be considered. The number of topologically 
distinct clusters with 1 through 8 triangles, and the sum of their lattice 
constant is shown in Table \ref{triangleclustersK}. (We have grouped them 
by the number of triangles.)
\begin{table}[h]
\caption{Kagom\'e lattice number of topological clusters and sum of the 
lattice constants for clusters with up to eight triangles.
The cluster with 0 triangles is the single site.}
\begin{tabular}{rrr}
\hline\hline
No.\ of triangles & No.\ of topological clusters & $\sum L(c)$ \\
\hline 
   0  &       1  &\qquad          1 \\
   1  &       1  &\qquad        2/3 \\
   2  &       1  &\qquad          1 \\
   3  &       1  &\qquad          2 \\
   4  &       2  &\qquad       14/3 \\
   5  &       2  &\qquad         12 \\
   6  &       5  &\qquad       94/3 \\
   7  &       7  &\qquad      250/3 \\
   8  &      15  &\qquad        225 \\
\hline\hline
\end{tabular}
\label{triangleclustersK}
\end{table}

In Ref.\ \cite{rigol06} we have already discussed extensively
several spin models on the kagom\'e lattice. Hence, here 
we will restrict our analysis to the specific heat and the uniform susceptibility of
the AFHM.

\subsection{Heisenberg model}

The kagom\'e-lattice AFHM model has been argued to have
short-ranged spin-spin correlations down to 
$T=0$ \cite{lecheminant97,zeng95,singh92}. 
Its thermodynamic properties have also been of 
considerable interest \cite{sindzingre00}. 
In particular, an issue that is still under debate (motivated by experiments 
on He$^3$ on graphite) is whether this model exhibits a two-peaked structure 
in the specific heat \cite{elser89,elstner94}.

\begin{figure}[!htb]
\begin{center}
  \includegraphics[scale=.6,angle=0]{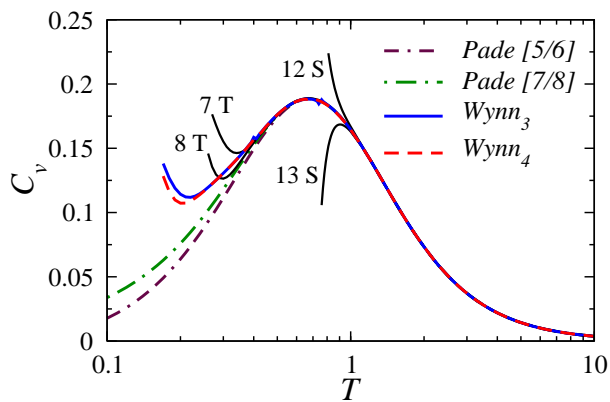}
\end{center}
\vspace{-0.5cm}
\caption{\label{EXTRAP_CvHeisenbergKA}
(Color online) 
Specific heat as a function of temperature for the Heisenberg model 
on a kagom\'e lattice. Direct sums, for up to 7 and 8 triangles 
(7 T and 8 T in the figure) and up to 12 and 13 sites 
(12 S and 13 S in the figure), are compared with extrapolations for
the triangle expansion and with two Pade approximants from 
Ref.\ \cite{elstner94}.}
\end{figure}

In Ref.\ \cite{rigol06} we have already studied the specific heat 
of the AFHM. There we compared the direct sums of the triangle expansion 
with Pade approximants from HTE \cite{elstner94}, which showed an overall 
good agreement down to $T\sim 0.3$. The triangle expansion for the specific
heat on the kagom\'e lattice, in contrast to the triangular lattice in 
the previous section, allows for an acceleration of the convergence by means 
of Wynn's extrapolations.

In Fig.\ \ref{EXTRAP_CvHeisenbergKA}, we compare
the bare sums for the site and triangle expansions with results of Wynn's
extrapolation and Pade approximants \cite{elstner94}. As seen in 
Fig.\ \ref{EXTRAP_CvHeisenbergKA}, the results for Wynn's extrapolation 
appear to converge down to $T\sim 0.2$ and exhibit a clear deviation from 
the Pade results. The deviations are, one could say, in the right direction
since the extrapolation of the specific-heat HTE down to $T=0$ has a large 
missing entropy \cite{elstner94}.

\begin{figure}[!htb]
\begin{center}
  \includegraphics[scale=.6,angle=0]{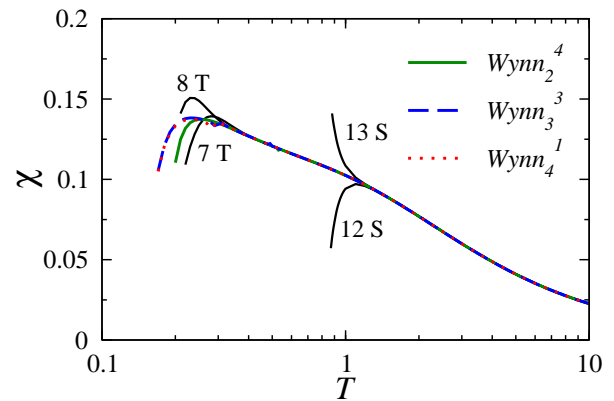}
\end{center}
\vspace{-0.5cm}
\caption{\label{EXTRAP_SusceptibilityHeisenbergKA}
(Color online) 
Uniform susceptibility as a function of temperature for the AFHM on a 
kagom\'e lattice. NLC bare sums are shown for up to 7 and 8 triangles 
(7 T and 8 T in the figure), and up to 12 and 13 sites (12 S and 13 S 
in the figure). The corresponding Wynn's extrapolations for triangle based 
expansions are also shown.}
\end{figure}

To conclude this section we show in Fig.\ \ref{EXTRAP_SusceptibilityHeisenbergKA} 
NLC results for the bare and extrapolated sums of the uniform susceptibility 
($\chi$) of the AFHM on the kagom\'e lattice. Similar to the extrapolations 
for the specific heat, Wynn's extrapolations are well converged down to $T\sim 0.2$, 
while the direct sums for the triangle expansion converge down to $T\sim 0.3$. 
On the other hand, the site based expansion, up to 13 sites, converges only down 
to $T\sim 1$, which is the same temperature one can access with HTE without 
extrapolations. Overall, for the kagom\'e lattice we have found that direct 
and extrapolated sums of the triangle based expansion converge better for 
the thermodynamic observables considered (energy, entropy, specific heat, 
and uniform susceptibility) than the site and bond based expansions.

\section{Conclusions \label{conclusions}}

We have presented an extensive discussion of the numerical linked 
cluster algorithm introduced in Ref.\ \cite{rigol06}. We have 
detailed how to construct NLC starting from different building blocks 
on square, triangular, and  kagom\'e lattices. Several sequence
extrapolation techniques, which we have used to accelerate NLC 
convergence, have also been discussed.

In order to show how NLC works for models with different ground
states and orders, we have studied several spin models on square,
triangular, and kagom\'e lattices. We have shown that NLC is
better suited for systems that remain short ranged at all temperatures
(such as the $XY$ model in a staggered field), and for models where correlations
build up slowly so that they become large only at very low temperatures.
A good example of the latter case is the AFHM on the kagom\'e lattice,
for which well converged results could be obtained down to $T\sim 0.3$ 
without the need of using sequence extrapolation techniques.

Similar to HTE, NLC also allows for extrapolations beyond the region of 
convergence provided by clusters up to a given system size. 
It is important to note that within NLC the region of convergence 
is dictated by the largest cluster sizes considered, and by
the range of correlations in the thermodynamic system.
Hence, even without extrapolations, one can, in principle, 
extend the region of convergence by including larger clusters. 
In this respect NLC is fundamentally different from HTE, whose region 
of convergence is dictated by the dominant microscopic energy scale, 
and including larger clusters can only help with extrapolations as
they do not improve the convergence of the direct sum. Extrapolations 
within NLC allow one to go to temperatures lower than accessible 
by means of direct Pade approximants for HTE. Examples 
where NLC is superior in this respect include the specific 
heat in the triangular and kagom\'e lattices.

Finally, we have also compared NLC results with those obtained from 
exact diagonalization of clusters with periodic boundary conditions.
We have shown that NLC provides accurate results even where ED still 
suffers from very large finite size effects. Even for short ranged 
models such as the $XY$ model in a staggered field, ED can fail to predict,
for example, the position and height of the peak in the specific heat.

Although it was not implemented here, one way to improve NLC 
convergence at lower temperatures would be to use Lanczos type methods to
diagonalize larger clusters. Larger clusters would become possible if one was to focus
only on low lying states rather than the full diagonalization that we have used in 
this work.

\begin{acknowledgments}

This work was supported by the US National Science Foundation, 
Grant Nos.\ DMR-0240918, DMR-0312261, and PHY-0301052. We are grateful 
to B. Bernu and G. Misguich for providing us with their data from 
Ref.\ \cite{bernu01}, and to R. Yu for providing us with the 
QMC results presented in Sec.\ \ref{SQ}.

\end{acknowledgments}

\end{document}